\documentclass[9pt,conference]{IEEEtran}
\IEEEoverridecommandlockouts
\usepackage{amsmath,amssymb,graphicx}
\usepackage{tikz}
\usetikzlibrary{arrows.meta,positioning,shapes.geometric,fit,backgrounds}
\usepackage{booktabs}
\usepackage{multirow}
\usepackage{indentfirst}
\usepackage{balance}
\usetikzlibrary{calc}
\usepackage{subcaption}
\usepackage[section]{placeins} 

\def\BibTeX{{\rm B\kern-.05em{\sc i\kern-.025em b}\kern-.08em
    T\kern-.1667em\lower.7ex\hbox{E}\kern-.125emX}}
\title{CASR: Content-Adaptive Neural Super-Resolution Post-Filter for Versatile Video Coding via Low-Rank Overfitting

\thanks{*Corresponding author: \texttt{khoa.phamdinh@aalto.fi}. Khoa Pham-Dinh was a student at Tampere University during part of the work.}

}

\author{
    Khoa Pham-Dinh\textsuperscript{*,1,2,3}, Francesco Cricri\textsuperscript{1}, Maria Santamaria\textsuperscript{1}, Honglei Zhang\textsuperscript{1},\\
    Hamed R. Tavakoli\textsuperscript{1,2}, Moncef Gabbouj\textsuperscript{3}, Juho Kannala\textsuperscript{2}, Miska M. Hannuksela\textsuperscript{1}\\
    \textsuperscript{1}\textit{Nokia}, \textsuperscript{2}\textit{Aalto University}, \textsuperscript{3}\textit{Tampere University}\\
}

\begin{document}

\maketitle

\begin{abstract}
The use of super-resolution as a post-processing step following a video codec allows videos to be encoded at reduced spatial resolution at the encoder side and to be reconstructed and upsampled to the original resolution at the decoder side. In this way, the required bitrate is reduced and the quality of the reconstructed frames is improved without modifying the core coding architecture. However, generic SR models are typically trained offline and lack sufficient adaptability to the diverse content characteristics and compression artifacts produced by video codecs, which limits their effectiveness during test time. To address the limitation, this paper proposes CASR (Content-Adaptive Super-Resolution), a content-adaptive SR post-filter framework for Versatile Video Coding (VVC), based on encoder-side overfitting on each input sequence. In order to limit the bitrate overhead required for signalling the content adaptation signal, i.e., the weight-update, Low-Rank Adaptation (LoRA) is leveraged. The method freezes the convolution kernels of a pretrained SR network and fine-tunes only lightweight rank-$r$ matrices attached to selected convolution layers, using VVC decoded frames and quantization-parameter (QP) maps of test sequences as supervision. The resulting low-rank update is compressed with the MPEG Neural Network Compression and Representation (NNR) standard. Experiments on the JVET common test conditions (CTC) class A1 and A2 sequences indicate that LoRA-based content adaptation provides bitrate savings over a non-adapted SR post-filter at a small signalling cost. Compared with the VVC Test Model (VTM~21.0), the proposed method achieves BD-rate savings of $-10.93\%$ (Y), $-15.39\%$ (U), $-24.41\%$ (V) under random access and $-13.43\%$ (Y), $-5.75\%$ (U), $-22.94\%$ (V) under all-intra. An ablation of the LoRA rank $r$ further shows that $r{=}4$ provides the best trade-off between coding gain and signalling cost.
\end{abstract}


\section{Introduction}
\label{sec:intro}
The latest video coding standard developed by the Joint Video Experts Team (JVET) is the Versatile Video Coding (VVC) standard, which was released in July 2020 \cite{brossOverviewVersatileVideo2021}. VVC achieves substantial compression efficiency gains over previous video coding standards, enabling significant bitrate reduction for a wide range of applications~\cite{hamidoucheVersatileVideoCoding2022}. Nevertheless, at low and medium bitrates, reconstructed video frequently suffers from compression artifacts and loss of fine spatial details due to aggressive quantization and block-based processing. Therefore, improving reconstruction quality without modifying standardized coding tools remains an important challenge. To reduce coding artifacts, VVC performs in-loop filtering, which is applied inside the prediction–reconstruction loop to reconstructed frames or blocks, so that the filtered signal is used as a reference for subsequent prediction. However, it is possible to perform post-processing filtering on the VVC-decoded pictures to reduce any remaining visible artifacts and improve perceptual quality without affecting prediction or bitstream syntax. In this paper, we focus on post-processing filtering techniques, with particular emphasis on super-resolution methods.

Decoder-side post-processing has emerged as an effective strategy to enhance reconstructed video quality while preserving full compliance with video coding standards. Previous work on post-filtering has been study such as Zhao et al.~\cite{zhaoCNNBasedPostProcessingAlgorithm2020} studies post processing filter on the High Efficiency Video Coding (HEVC)~\cite{hevc2012} standard; Zhang et al.~\cite{zhangEnhancingVVCCnnBased2020} proposed CNN-based post-processing for VVC using input RGB and quantization parameter (QP), the method employs five separate QP-specific trained CNN models for adaptive post-processing; Zhang et al.~\cite{zhangWCDANNLightweightCNN2023} introduce a lightweight post-processing filter that uses depthwise separable convolution and dual attention for VVC artifact removal. However, most existing post-filtering approaches rely on neural networks trained offline on generic datasets and applied uniformly across all content. To address the limited adaptability of generic neural post-filters, prior works have investigated content-adaptive neural network techniques, starting with Lam et al.~\cite{lamCompressingWeightupdatesImage2019} introduce a weight-update artifact removal filter, Santamaria et al.~\cite{santamariaContentadaptiveConvolutionalNeural2021}~\cite{santamariaOverfittingMultiplierParameters2022} proposed content adaptive scheme updating only selected bias or multiplier based on energy to require much lower bitrate overhead for transmitting weight-updates. Moreover, to effectively compress the weight update for use in standardization activities, the MPEG Neural Network Compression and Representation (NNR) Standard ~\cite{kirchhofferOverviewNeuralNetwork2022} was proposed as a comprehensive toolbox combining pre-processing methods (sparsification, pruning, low-rank decomposition), quantization techniques, and DeepCABAC entropy coding~\cite{wiedemannDeepCABACUniversalCompression2020} to achieve up to $97\%$ compression efficiency while maintaining inference accuracy. These approaches improve performance by specializing the filtering network to the target video sequence. Recent standardization efforts have addressed the signalling of neural network–based enhancement tools through the Versatile Supplemental Enhancement Information (VSEI) standard (ISO/IEC 23002-7, ITU-T H.274)~\cite{H274VersatileSupplemental}. VSEI defines optional Supplemental Enhancement Information (SEI) messages that allow neural network post-filters and associated parameter updates to be conveyed to the decoder without modifying the core decoding process. In particular, VSEI introduces dedicated SEI messages for neural network post-filter (NNPF) to describe the characteristics of the post-filter neural network and to control their activation on a picture basis, enabling backward-compatible deployment of AI-based post-processing in systems using VVC, HEVC, or Advanced Video Coding (AVC)~\cite{avc2003} standards.

Among post-processing approaches, super-resolution (SR) has received considerable attention, as it enables video content to be encoded at reduced spatial resolution and subsequently reconstructed at higher resolution after decoding, leading to improved coding efficiency and visual quality~\cite{khaniEfficientVideoCompression2021}. Super-resolution has been explored both as a standalone post-processing tool and as part of hybrid video coding pipelines. For example, Nvidia’s Deep Learning Super Sampling (DLSS) technology relies on neural super-resolution to upscale lower-resolution frames to a target display resolution~\cite{nvidiacorporationDLSS4Transforming2025}. Super-resolution approaches have also been studied for post filtering of HEVC bitstreams~\cite{wangEfficientSuperResolutionCompression2023}. In addition, Reference Picture Resampling (RPR) in VVC enables resolution changes during encoding to adapt to varying network conditions without inserting intra pictures~\cite{fuEfficientFrameworkReference2022}. Several works build upon RPR for neural-network-based super-resolution (NNSR) \cite{linLowComplexitySuper2025} \cite{huangCNNFilterSuperResolution2023} and have been adopted into the JVET NNVC (Neural-Network-based Video Coding) exploration model \cite{lvEE141MultipleScaling2024}. However, the NNSR approach in NNVC works as an enhancement filter, where the resolution change is performed by the RPR upsampler and the neural network only refines the resampled pictures. This limits the applicability when super-resolution is desired as an independent decoder-side post-filter. In particular, to remain compatible with the VSEI NNPF signalling, a post-processing filter with the purpose of super-resolution should perform the upsampling operation within the post-processing model itself, rather than relying on the non-normative use of the RPR upsampler.

In this paper, we propose CASR, a content-adaptive SR post-filtering framework for VVC, where the model's interface (e.g., input and output tensors) is compliant with the VSEI NNPF signalling. The base SR network performs $\times 2$ upscaling using sub-pixel convolution~\cite{shiSubpixelConvolution2016} and is adapted from the JVET NNSR architecture in~\cite{lvEE141MultipleScaling2024}, which we modify to (i) include the upsampling operation inside the post-filter and (ii) accept the inputs and signalling required by the VSEI NNPF SEI messages. For content adaptation, we depart from previous multiplier-only schemes~\cite{santamariaOverfittingMultiplierParameters2022} and instead apply Low-Rank Adaptation (LoRA)~\cite{huLoRALowRank2022} to selected convolution layers of the SR network. The convolution weights of the pretrained network are frozen, and only two low-rank matrices per layer are fine-tuned at the encoder side using the VVC-decoded test frames as supervision. The resulting low-rank update is encoded with the NNR toolbox~\cite{kirchhofferOverviewNeuralNetwork2022} using an adaptive QP search and DeepCABAC~\cite{wiedemannDeepCABACUniversalCompression2020}, and is transmitted to the decoder. The contributions of this paper are: (1) an SR post-filter model for VVC that is compatible with the VSEI standard, both in its input/output tensors and in incorporating the upsampling operation as part of the network; (2) a LoRA-based content-adaptation scheme that reduces the number of trainable per-sequence parameters compared with full fine-tuning while preserving full compatibility with the VVC decoder; and (3) an evaluation on the JVET CTC class A test set shows the trade-off between pretrained-only SR post-filtering, multiplier, and the proposed LoRA adaptation.

The remainder of the paper is organized as follows. Section~\ref{sec:methodology} describes our SR models and content-adaptive overfitting, Section~\ref{sec:result} reports and discusses the results, and Section~\ref{sec:conclusion} concludes the paper.

\section{Methodology}
\label{sec:methodology}
\begin{figure*}[t]
    \centering
    \includegraphics[width=1.4\columnwidth]{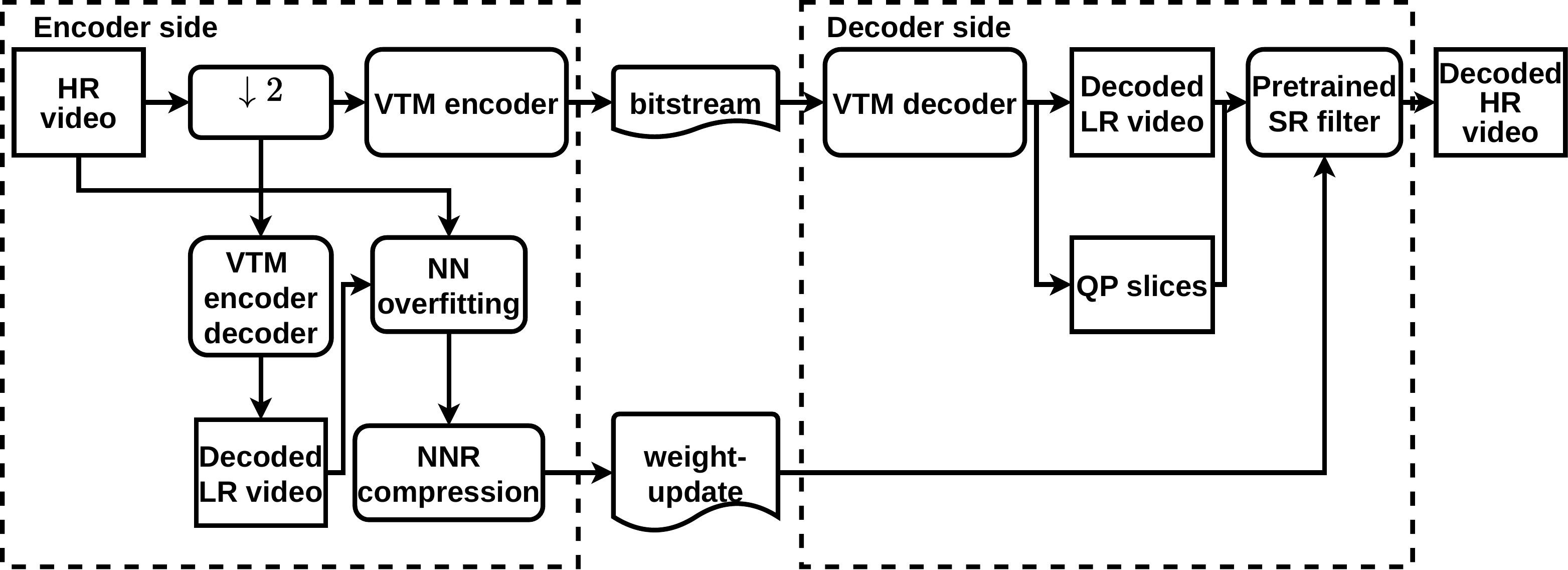}
    \caption{Proposed VSEI-compliant SR post-filtering pipeline with encoder side content adaptation.}
    \label{fig:2_pipeline}
\end{figure*}

Figure~\ref{fig:2_pipeline} illustrates the proposed framework. The input high-resolution (HR) sequence is downsampled by a factor $s=2$ in both spatial dimensions and encoded with VVC Test Model (VTM~21.0) under the random access (RA) and all-intra (AI) configurations. At the decoder, the decoded low-resolution (LR) frames are enhanced by an SR post-filter that is fully outside the coding loop. The post-filter is overfitted to each sequence at the encoder side by training only the LoRA matrices or the multiplier matrices attached to each convolution while the pretrained convolution kernels remain frozen. The resulting update is quantised, entropy-encoded with NNR~\cite{kirchhofferOverviewNeuralNetwork2022}, and transmitted as side information. At the decoder, the NNR bitstream is decoded, the weight update is merged to the frozen base weight, and the adapted SR network performs upsampling.
\subsection{Network architecture}
\label{sec:sr_arch}
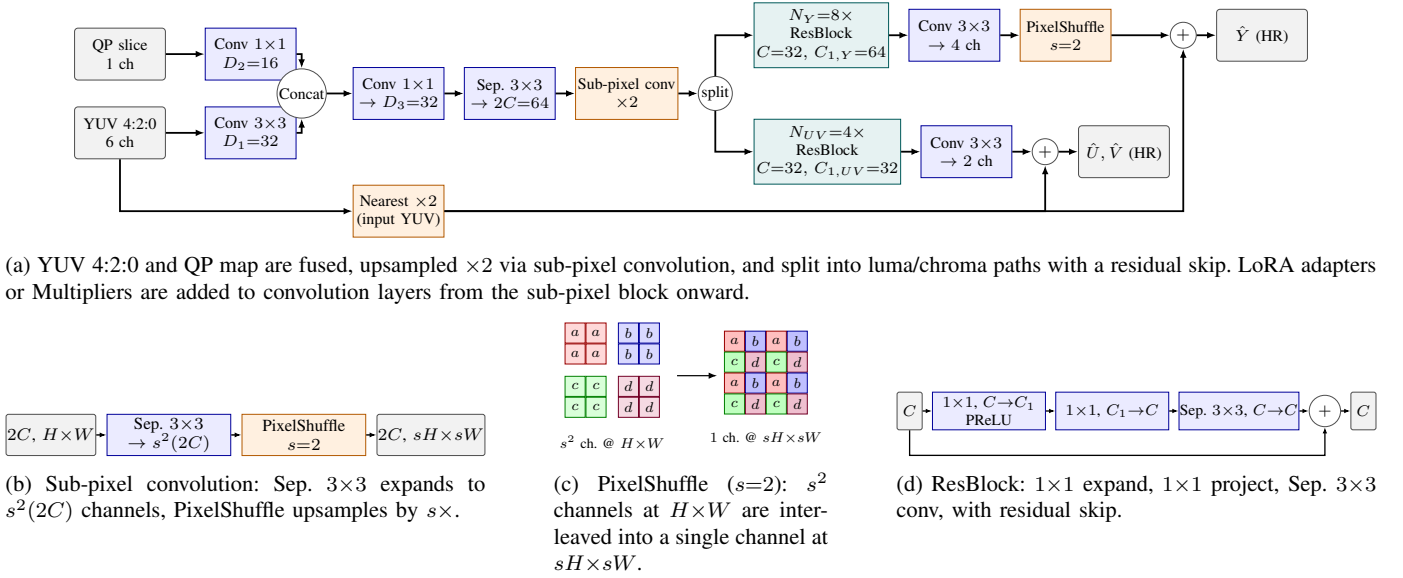
\begin{figure*}[t]
\centering

\begin{subfigure}{\textwidth}
\centering
\resizebox{0.9\textwidth}{!}{%
\begin{tikzpicture}[
    font=\scriptsize,
    every node/.style={align=center},
    arr/.style={-{Latex[length=1.4mm]}, thick},
    skip/.style={-{Latex[length=1.4mm]}, thick},
    io/.style={rectangle, rounded corners=1.5pt, draw=black!70, fill=black!5,
               minimum width=14mm, minimum height=8mm, inner sep=1pt},
    conv/.style={rectangle, draw=blue!60!black, fill=blue!8,
                 minimum width=14mm, minimum height=8mm, inner sep=1pt},
    ps/.style={rectangle, draw=orange!70!black, fill=orange!12,
               minimum width=14mm, minimum height=8mm, inner sep=1pt},
    rb/.style={rectangle, draw=teal!70!black, fill=teal!10,
               minimum width=18mm, minimum height=10mm, inner sep=1pt},
    op/.style={circle, draw=black!70, fill=white,
               inner sep=0.5pt, minimum size=4mm},
]
\node[io]                       (yuv)   {YUV 4:2:0\\6 ch};
\node[io, above=4mm of yuv]     (qp)    {QP slice\\1 ch};
\node[conv, right=6mm of yuv]   (cyuv)  {Conv $3{\times}3$\\$D_1{=}32$};
\node[conv, right=6mm of qp]    (cqp)   {Conv $1{\times}1$\\$D_2{=}16$};
\draw[arr] (yuv) -- (cyuv);
\draw[arr] (qp)  -- (cqp);
\node[op, right=4mm of $(cyuv)!0.5!(cqp)$] (cat) {Concat};
\draw[arr] (cyuv.east) -| (cat.south);
\draw[arr] (cqp.east)  -| (cat.north);
\node[conv, right=4mm of cat]   (merge) {Conv $1{\times}1$\\$\to D_3{=}32$};
\draw[arr] (cat)   -- (merge);
\node[conv, right=3mm of merge] (sep1)  {Sep.\ $3{\times}3$\\$\to 2C{=}64$};
\draw[arr] (merge) -- (sep1);
\node[ps, right=3mm of sep1]    (spc)   {Sub-pixel conv\\$\times 2$};
\draw[arr] (sep1) -- (spc);
\node[op, right=3mm of spc]     (split) {split};
\draw[arr] (spc) -- (split);
\node[rb, above right=2mm and 4mm of split] (yblk)
    {$N_Y{=}8{\times}$\\ResBlock\\$C{=}32,\,C_{1,Y}{=}64$};
\node[rb, below right=2mm and 4mm of split] (uvblk)
    {$N_{UV}{=}4{\times}$\\ResBlock\\$C{=}32,\,C_{1,UV}{=}32$};
\draw[arr] (split.north) |- (yblk.west);
\draw[arr] (split.south) |- (uvblk.west);
\node[conv, right=3mm of yblk]  (yout)  {Conv $3{\times}3$\\$\to 4$ ch};
\node[conv, right=3mm of uvblk] (uvout) {Conv $3{\times}3$\\$\to 2$ ch};
\draw[arr] (yblk)  -- (yout);
\draw[arr] (uvblk) -- (uvout);
\node[ps, right=3mm of yout]    (psy)   {PixelShuffle\\$s{=}2$};
\draw[arr] (yout) -- (psy);
\node[op, right=9mm of psy]     (sumy)  {$+$};
\node[op, right=3mm of uvout]   (sumuv) {$+$};
\draw[arr] (psy)   -- (sumy);
\draw[arr] (uvout) -- (sumuv);
\node[io, right=3mm of sumy]    (yhr)   {$\hat{Y}$ (HR)};
\node[io, right=3mm of sumuv]   (uvhr)  {$\hat{U},\hat{V}$ (HR)};
\draw[arr] (sumy)  -- (yhr);
\draw[arr] (sumuv) -- (uvhr);
\node[ps, below=10mm of merge]  (up)    {Nearest $\times 2$\\(input YUV)};
\draw[skip] (yuv.south) |- (up.west);
\draw[skip] (up.east)   -| (sumy.south);
\draw[skip] (up.east)   -| (sumuv.south);
\end{tikzpicture}%
}
\caption{YUV~4:2:0 and QP map are fused, upsampled $\times2$ via
         sub-pixel convolution, and split into luma/chroma paths
         with a residual skip. LoRA adapters or Multipliers are added to
         convolution layers from the sub-pixel block onward.}
\label{fig:arch_a}
\end{subfigure}

\medskip

\begin{subfigure}[t]{0.35\textwidth}
\centering
\resizebox{\linewidth}{!}{%
\begin{tikzpicture}[font=\scriptsize, node distance=1.5mm,
    every node/.style={align=center},
    arr/.style={-{Latex[length=1mm]}, semithick},
    io/.style={rectangle, rounded corners=1pt, draw=black!70, fill=black!5,
               minimum width=8mm, minimum height=6mm, inner sep=0.4pt},
    conv/.style={rectangle, draw=blue!60!black, fill=blue!8,
                 minimum width=18mm, minimum height=6mm, inner sep=0.4pt},
    ps/.style={rectangle, draw=orange!70!black, fill=orange!12,
               minimum width=18mm, minimum height=6mm, inner sep=0.4pt},
]
\node[io]                  (in)  {$2C$, $H{\times}W$};
\node[conv, right=of in]   (k1)  {Sep.\ $3{\times}3$\\$\to s^{2}(2C)$};
\node[ps,   right=of k1]   (ps)  {PixelShuffle\\$s{=}2$};
\node[io,   right=of ps]   (out) {$2C$, $sH{\times}sW$};
\draw[arr] (in) -- (k1);
\draw[arr] (k1) -- (ps);
\draw[arr] (ps) -- (out);
\end{tikzpicture}%
}
\caption{Sub-pixel convolution: Sep.\ $3{\times}3$ expands to
         $s^{2}(2C)$ channels, PixelShuffle upsamples by $s{\times}$.}
\label{fig:arch_b}
\end{subfigure}%
\hfill
\begin{subfigure}[t]{0.2\textwidth}
\centering
\resizebox{\linewidth}{!}{%
\begin{tikzpicture}[font=\scriptsize, node distance=0pt,
    arr/.style={-{Latex[length=1.5mm]}, semithick},
    cellA/.style={draw=red!60!black,    fill=red!15,    minimum size=3.5mm,
                  inner sep=0pt, font=\scriptsize},
    cellB/.style={draw=blue!60!black,   fill=blue!15,   minimum size=3.5mm,
                  inner sep=0pt, font=\scriptsize},
    cellC/.style={draw=green!50!black,  fill=green!15,  minimum size=3.5mm,
                  inner sep=0pt, font=\scriptsize},
    cellD/.style={draw=purple!60!black, fill=purple!15, minimum size=3.5mm,
                  inner sep=0pt, font=\scriptsize},
    cellAo/.style={draw=red!60!black,   fill=red!25,    minimum size=3.5mm,
                   inner sep=0pt, font=\scriptsize},
    cellBo/.style={draw=blue!60!black,  fill=blue!25,   minimum size=3.5mm,
                   inner sep=0pt, font=\scriptsize},
    cellCo/.style={draw=green!50!black, fill=green!25,  minimum size=3.5mm,
                   inner sep=0pt, font=\scriptsize},
    cellDo/.style={draw=purple!60!black,fill=purple!25, minimum size=3.5mm,
                   inner sep=0pt, font=\scriptsize},
]
\node[cellA] (c1r1c1) {$a$};
\node[cellA, right=0pt of c1r1c1] (c1r1c2) {$a$};
\node[cellA, below=0pt of c1r1c1] (c1r2c1) {$a$};
\node[cellA, right=0pt of c1r2c1] (c1r2c2) {$a$};

\node[cellB, right=2mm of c1r1c2] (c2r1c1) {$b$};
\node[cellB, right=0pt of c2r1c1] (c2r1c2) {$b$};
\node[cellB, below=0pt of c2r1c1] (c2r2c1) {$b$};
\node[cellB, right=0pt of c2r2c1] (c2r2c2) {$b$};

\node[cellC, below=2mm of c1r2c1] (c3r1c1) {$c$};
\node[cellC, right=0pt of c3r1c1] (c3r1c2) {$c$};
\node[cellC, below=0pt of c3r1c1] (c3r2c1) {$c$};
\node[cellC, right=0pt of c3r2c1] (c3r2c2) {$c$};

\node[cellD, right=2mm of c3r1c2] (c4r1c1) {$d$};
\node[cellD, right=0pt of c4r1c1] (c4r1c2) {$d$};
\node[cellD, below=0pt of c4r1c1] (c4r2c1) {$d$};
\node[cellD, right=0pt of c4r2c1] (c4r2c2) {$d$};

\node[font=\scriptsize,
      below=2mm of $(c3r2c1.south)!0.5!(c4r2c2.south)$, anchor=north]
    {$s^{2}$ ch.\ @ $H{\times}W$};

\draw[arr] ($(c4r2c2.east)+(3mm,5.5mm)$) -- ++(7mm,0);

\node[cellAo] (r1c1) at ($(c4r2c2.east)+(13mm,11.5mm)$) {$a$};
\node[cellBo, right=0pt of r1c1] (r1c2) {$b$};
\node[cellAo, right=0pt of r1c2] (r1c3) {$a$};
\node[cellBo, right=0pt of r1c3] (r1c4) {$b$};
\node[cellCo, below=0pt of r1c1] (r2c1) {$c$};
\node[cellDo, right=0pt of r2c1] (r2c2) {$d$};
\node[cellCo, right=0pt of r2c2] (r2c3) {$c$};
\node[cellDo, right=0pt of r2c3] (r2c4) {$d$};
\node[cellAo, below=0pt of r2c1] (r3c1) {$a$};
\node[cellBo, right=0pt of r3c1] (r3c2) {$b$};
\node[cellAo, right=0pt of r3c2] (r3c3) {$a$};
\node[cellBo, right=0pt of r3c3] (r3c4) {$b$};
\node[cellCo, below=0pt of r3c1] (r4c1) {$c$};
\node[cellDo, right=0pt of r4c1] (r4c2) {$d$};
\node[cellCo, right=0pt of r4c2] (r4c3) {$c$};
\node[cellDo, right=0pt of r4c3] (r4c4) {$d$};

\node[font=\scriptsize,
      below=2mm of $(r4c2.south)!0.5!(r4c3.south)$, anchor=north]
    {$1$ ch.\ @ $sH{\times}sW$};
\end{tikzpicture}%
}
\caption{PixelShuffle ($s{=}2$): $s^2$ channels at $H{\times}W$
         are interleaved into a single channel at $sH{\times}sW$.}
\label{fig:arch_c}
\end{subfigure}%
\hfill
\begin{subfigure}[t]{0.35\textwidth}
\centering
\resizebox{\linewidth}{!}{%
\begin{tikzpicture}[font=\scriptsize, node distance=1.5mm,
    every node/.style={align=center},
    arr/.style={-{Latex[length=1mm]}, semithick},
    skip/.style={-{Latex[length=1mm]}, semithick},
    io/.style={rectangle, rounded corners=1pt, draw=black!70, fill=black!5,
               minimum width=4mm, minimum height=6mm, inner sep=0.4pt},
    conv/.style={rectangle, draw=blue!60!black, fill=blue!8,
                 minimum width=18mm, minimum height=6mm, inner sep=0.4pt},
    op/.style={circle, draw=black!70, fill=white,
               inner sep=0pt, minimum size=5mm},
]
\node[io]                  (in)  {$C$};
\node[conv, right=of in]   (c1)  {$1{\times}1$, $C{\to}C_1$\\PReLU};
\node[conv, right=of c1]   (c2)  {$1{\times}1$, $C_1{\to}C$};
\node[conv, right=of c2]   (c3)  {Sep.\ $3{\times}3$, $C{\to}C$};
\node[op,   right=of c3]   (sum) {$+$};
\node[io,   right=of sum]  (out) {$C$};
\draw[arr] (in)  -- (c1);
\draw[arr] (c1)  -- (c2);
\draw[arr] (c2)  -- (c3);
\draw[arr] (c3)  -- (sum);
\draw[arr] (sum) -- (out);
\draw[skip] (in.south) -- ++(0,-4mm) -| (sum.south);
\end{tikzpicture}%
}
\caption{ResBlock: $1{\times}1$ expand, $1{\times}1$ project, Sep.\ $3{\times}3$ conv, with residual skip.}
\label{fig:arch_d}
\end{subfigure}

\caption{Network architecture: (a) full pipeline, (b) sub-pixel
         convolution block, (c) PixelShuffle rearrangement, (d) ResBlock.}
\label{fig:base_model}
\end{figure*}

The base SR network performs $\times 2$ upscaling from a decoded YUV 4:2:0 frame and a per-slice QP map. It follows a residual learning paradigm: the LR input $\mathbf{x}=(\mathbf{x}_{\mathrm{YUV}},\mathbf{x}_{\mathrm{QP}})$ is mapped to an HR residual that is added to a deterministic upsampling of the decoded picture,
\begin{equation}
\hat{\mathbf{y}} = \mathcal{U}_{\mathrm{NN}}(\mathbf{x}_{\mathrm{YUV}}) + \mathcal{S}(\mathbf{x};\boldsymbol{\theta}),
\label{eq:residual}
\end{equation}
where $\mathcal{U}_{\mathrm{NN}}$ is the nearest neighbour $\times 2$ upsampler and $\mathcal{S}(\cdot;\boldsymbol{\theta})$ is the trained residual network that includes an upsampling layer.

Figure~\ref{fig:base_model} shows the SR network architecture. In general, the base SR model requires 79.4kMACs per pixel and includes $137,091$ trainable parameters. The decoded YUV 4:2:0 frame is packed into a 6-channel LR tensor (4 sub-sampled luma channels via pixel-unshuffle, plus the U and V chroma channels). This tensor and the QP map are processed by two parallel input convolutions (3$\times$3 and 1$\times$1, $D_1{=}32$ and $D_2{=}16$ feature channels), concatenated along the channel dimension and merged by a 1$\times$1 convolution to $D_3{=}32$ channels. A separable 3$\times$3 convolution then projects the feature map to $2C{=}64$ channels, immediately followed by a sub-pixel convolution layer~\cite{shiSubpixelConvolution2016} that expands it to $s^2\cdot 2C{=}256$ channels, performs the $\times 2$ upscaling, and reduces the channel count back to $2C{=}64$. The HR feature map is split along the channel dimension into a luma path ($N_Y{=}8$ residual blocks of width $C{=}32$, expansion $C_{1,Y}{=}64$) and a chroma path ($N_{UV}{=}4$ residual blocks of width $C{=}32$, expansion $C_{1,UV}{=}32$). Each residual block is composed of a 1$\times$1 expansion (with PReLU), a 1$\times$1 projection, and a separable 3$\times$3 convolution, with a residual skip~\cite{lvEE141MultipleScaling2024}. Finally, two output convolutions produce a 4-channel luma residual (re-assembled by PixelShuffle into one HR luma plane) and a 2-channel chroma residual at HR. All nonlinearities are PReLU and inference is performed in fp32. Trainable LoRA adapters or multipliers are applied to convolution layers from the sub-pixel block onward.

\subsection{Content adaptation}
\label{sec:lora_overfit}

\newcommand{\caFigScale}{0.7}

\begin{figure}[t]
    \centering
    \includegraphics[width=\caFigScale\columnwidth]{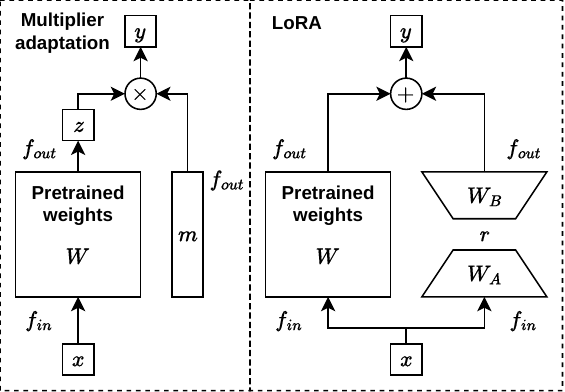}
    \caption{(a)~Channel-wise multiplier scaling and (b)~low-rank weight update (LoRA) attached to a frozen convolution layer.}
    \label{fig:content_adapt}
\end{figure}

\paragraph*{Multiplier baseline} As a reference adaptation scheme, we adopt the channel-wise multiplier overfitting of Santamaria et al.~\cite{santamariaOverfittingMultiplierParameters2022}. Each convolution block is augmented with a channel-wise learnable  multiplier vector $\mathbf{m}^{(l)}\in\mathbb{R}^{C_{\mathrm{out}}^{(l)}}$ that scales the post-activation output. For layer $l$ and output channel $i$,
\begin{equation}
y_i^{(l)} = \sigma\!\left(\mathbf{W}_i^{(l)} * \mathbf{x}^{(l)} + b_i^{(l)}\right)\cdot m_i^{(l)},
\label{eq:mult}
\end{equation}
where $*$ is convolution, $\mathbf{W}_i^{(l)}$ and $b_i^{(l)}$ are the frozen kernel and bias of channel $i$, and $\sigma(\cdot)$ is the PReLU non-linearity. All multipliers are initialised to $1$, so the network reproduces the pretrained behaviour at the start of overfitting. During encoder-side fine-tuning, only $\{m_i^{(l)}\}$ are trainable; the convolution kernels, biases, and PReLU coefficients are frozen.

\paragraph*{Proposed LoRA scheme} We model the per-sequence content adaptation as a low-rank update~\cite{huLoRALowRank2022} at selected convolution kernels of the base network. Let $\mathbf{W}^{(l)}\in\mathbb{R}^{C_{\mathrm{out}}^{(l)}\times C_{\mathrm{in}}^{(l)}\times k_h^{(l)}\times k_w^{(l)}}$ be the kernel of layer $l$, with $f_{\mathrm{in}}^{(l)}=C_{\mathrm{in}}^{(l)} k_h^{(l)} k_w^{(l)}$ and $f_{\mathrm{out}}^{(l)}=C_{\mathrm{out}}^{(l)}$. We attach two trainable matrices $A^{(l)}\in\mathbb{R}^{r\times f_{\mathrm{in}}^{(l)}}$ and $B^{(l)}\in\mathbb{R}^{f_{\mathrm{out}}^{(l)}\times r}$ and, at inference time, we replace the kernel by
\begin{equation}
\widetilde{\mathbf{W}}^{(l)} = \mathbf{W}^{(l)} + \frac{\alpha}{r}\,\mathrm{reshape}\!\left(B^{(l)} A^{(l)},\;\mathbf{W}^{(l)}\text{shape}\right),
\label{eq:lora}
\end{equation}
where $r\ll\min(f_{\mathrm{in}}^{(l)},f_{\mathrm{out}}^{(l)})$ is the LoRA rank and $\alpha$ a scaling factor. We initialise $A^{(l)}$ with Kaiming uniform and $B^{(l)}$ with zeros, so that $\widetilde{\mathbf{W}}^{(l)}{=}\mathbf{W}^{(l)}$ at the start of overfitting. During content adaptation only $\{A^{(l)}, B^{(l)}\}$ are trainable; the convolution kernels $\mathbf{W}^{(l)}$, biases, and PReLU coefficients of the pretrained network are frozen. This formulation generalises the channel-wise multiplier overfitting of~\cite{santamariaOverfittingMultiplierParameters2022}: a multiplier corresponds to a rank-1 update of a 1$\times$1 kernel, whereas LoRA allows a rank-$r$ update of every kernel shape (including the separable 3$\times$1 and 1$\times$3 layers used in our network) and shares parameters across channels. The number of trainable LoRA parameters per layer is $r(f_{\mathrm{in}}^{(l)}+f_{\mathrm{out}}^{(l)})$. Summed over all convolution layers of the network, this remains a small fraction of the base model parameters (reported per sequence in Sec.~\ref{sec:result}).

At the encoder, the LoRA matrices are fine-tuned for each sequence using the original HR frames as supervision. For each training tuple $(\mathbf{x}_n,\mathrm{qp}_n,\mathbf{y}_n)$ (decoded LR frame, per-slice QP map, original HR frame) we minimise a per-component MSE loss,
\begin{equation}
\mathcal{L} = \frac{w_Y\,\mathrm{MSE}\!\left(\hat{\mathbf{y}}_Y,\mathbf{y}_Y\right) + w_{UV}\,\mathrm{MSE}\!\left(\hat{\mathbf{y}}_{UV},\mathbf{y}_{UV}\right)}{w_Y + w_{UV}},
\label{eq:loss}
\end{equation}
with $w_Y=4$, $w_{UV}=2$, computed on the 4-channel luma residual stream and the 2-channel chroma residual stream respectively. The same loss function is used for both base-model pretraining and per-sequence content adaptation. We use multiplier-only overfitting~\cite{santamariaOverfittingMultiplierParameters2022} as a comparable method for comparison. Figure~\ref{fig:content_adapt} illustrates both adaptation methods applied to a frozen convolution layer.

\subsection{NNR coding}
\label{sec:update_coding}
Both methods produce a per-sequence weight update encoded with the NNR toolbox~\cite{kirchhofferOverviewNeuralNetwork2022}: $\Delta\boldsymbol{\theta}=\{A^{(l)},B^{(l)}\}_{l}$ for LoRA and $\Delta\boldsymbol{\theta}=\{m^{(l)}\}_{l}$ for the Multiplier. Uniform-scalar quantisation (without dependent quantisation) is followed by DeepCABAC entropy coding~\cite{wiedemannDeepCABACUniversalCompression2020}. The NNR QP is \emph{adaptive}: from an initial value of $-40$, an inner rate--distortion search refines the per-tensor QP to minimise bitstream length subject to a tolerated reconstruction-loss change on the encoder-side training loader. The selected QP is signalled in the bitstream payload. At the decoder, $\Delta\boldsymbol{\theta}$ is dequantised and merged into the frozen base weights once per sequence via Eq.~\eqref{eq:mult} or Eq.~\eqref{eq:lora}.

\section{Results and discussion}
\label{sec:result}

\subsection{Experimental setup}
\label{sec:setup}
\textbf{Base-model pretraining.}
The base SR network is trained on DIV2K~\cite{agustssonNTIRE2017Challenge2017} for the all-intra (AI) configuration and on the UHD subset of BVI-DVC~\cite{maBVIDVCTrainingDatabase2022} and TVD~\cite{xuTencentVideoDataset2021} for the random-access (RA) configuration. Corresponding validation set of these test set are used for validation. Training uses Adam (lr$\,{=}\,1\mathrm{e}{-}3$) with ReduceOnPlateau scheduling (patience 10, factor 0.5, minimum lr$\,{=}\,1\mathrm{e}{-}6$) for up to 200 epochs and batch size 256.

\textbf{Content-adaptive overfitting.}
For each test sequence, only the content-adaptive parameters (LoRA matrices or multiplier scalars) are fine-tuned while the base-model weights remain frozen. Overfitting uses Adam (lr$\,{=}\,1\mathrm{e}{-}3$) with the same ReduceOnPlateau schedule (patience 10, factor 0.5, minimum lr$\,{=}\,1\mathrm{e}{-}6$) for up to 200 epochs, batch size 16, $64{\times}64$ LR patches. LoRA uses default rank $r{=}4$ and $\alpha{=}4$.

\textbf{Evaluation.}
Experiments are conducted on the JVET CTC Class~A1 (Tango2, FoodMarket4, Campfire) and Class~A2 (CatRobot1, DaylightRoad2, ParkRunning3) sequences at $3840{\times}2160$. Each sequence is downsampled by ${\times}2$ before encoding so that the SR post-filter restores the native HR resolution. Encoding uses VTM under both RA and AI configurations with five QPs $\{22,\, 27,\, 32,\, 37,\, 42\}$. The anchor is VTM compression at full resolution with five QPs $\{27,\, 32,\, 37,\, 42,\, 47\}$. The mismatch in QP is intentional to ensure that bitrates are matched, since the proposed methods encode at half resolution, resulting in significantly lower bitrate at the same QP. BD-rate is computed using the Bj{\o}ntegaard Delta formulation~\cite{barmanBjontegaardDeltaBD2024} with native-resolution VTM as the anchor. For all content-adaptive methods, the bitrate includes the NNR-coded side information added to the VVC bitstream. SR inference is performed in fp32 PyTorch.

\subsection{Overall Results and Analysis}
\begin{table}[t]
\caption{BD-rate (\%) and relative encoding/decoding time (EncT/DecT)
compared to VTM. While both overfitting methods improve the base model,
LoRA significantly outperforms Multiplier on chroma (UV) components.}
\label{tab:main_results}
\centering
\footnotesize
\setlength{\tabcolsep}{3pt}
\renewcommand{\arraystretch}{0.8}
\begin{tabular}{l|rrr|rr}
\toprule
 & \multicolumn{3}{c|}{RA} & & \\
Method          & Y (\%)  & U (\%)  & V (\%) & EncT & DecT \\
\midrule
RPR             & -3.94   & 6.62    & 4.83   & 75\%  & 67\%  \\
Base SR         & -1.76   & 34.63   & 19.94  & 74\%  & 81\%  \\
Multiplier (approx. $100\%$ EncT)   & -9.41 &-2.75&	-5.94  & 102\% & 95\%  \\
Multiplier      & -9.50   & -3.40   & -6.76  & 147\% & 95\%  \\
LoRA (approx. $100\%$ EncT)      & -10.46   &	-14.08    &	-22.88  & 101\% & 82\%  \\
LoRA            & \textbf{-10.93}  & \textbf{-15.39}  & \textbf{-24.41} & 145\% & 82\%  \\
\midrule
 & \multicolumn{3}{c|}{AI} & & \\
Method          & Y (\%)  & U (\%)  & V (\%) & EncT & DecT \\
\midrule
RPR             & -4.04   & 16.65   & 14.76  & 90\%  & 46\%  \\
Base SR         & -5.08   & 37.61   & 13.76  & 70\%  & 59\%  \\
Multiplier (approx. $100\%$ EncT)       & -12.07   & 9.55   & -5.20  & 102\% & 69\%  \\
Multiplier      & -12.26  & 7.77    & -6.76  & 176\% & 69\%  \\
LoRA (approx. $100\%$ EncT)      & -12.86   & -3.36   & -20.54  & 98\% & 59\%  \\
LoRA            & \textbf{-13.43}  & \textbf{-5.75}   & \textbf{-22.94} & 170\% & 59\%  \\
\bottomrule
\end{tabular}
\end{table}

Table~\ref{tab:main_results} summarizes the BD-rate results under both RA and AI configurations. The RPR baseline uses the RPR upsampling filter of VVC as a non-normative post-filter and achieves moderate luma savings ($-3.94\,\%$ for RA and $-4.04\,\%$ for AI) but degrades chroma quality, confirming that resolution adaptation alone cannot preserve colour fidelity. The non-adapted Base SR model exhibits a similar pattern more acutely: while it provides luma gains, its chroma BD-rates are severely degraded. Both RPR and Base SR results show that super-resolution alone fails to recover colour. In contrast, the overfitting approaches deliver substantial luma improvements, with Multiplier reaching $-9.50\,\%$ and $-12.26\,\%$ and LoRA reaching $-10.93\,\%$ and $-13.43\,\%$ for Y under RA and AI, respectively, while also mitigating the chroma deficiencies. 

In particular, the Multiplier approach improves significantly over Base SR, turning most chroma BD-rates negative for RA. However, some residual positive values remain for AI (e.g.\ $+7.77\,\%$ for U under AI), indicating that scalar gain adjustment has limited capacity to capture spatially varying colour patterns. LoRA adaptation, by contrast, achieves consistent savings across all three components in both configurations, reaching $-15.39\,\%$/$-24.41\,\%$ for U/V under RA. The improvement over Multiplier is particularly pronounced on chroma, confirming that low-rank weight updates offer a richer adaptation space than element-wise scaling.

In terms of complexity, both adaptation methods increase encoding time to 145--176\,\% of VTM due to the iterative fine-tuning loop. On the decoding side, the two methods differ slightly: LoRA's decoding time is identical to that of the Base SR, since its low-rank weights are merged into the model parameters at inference and add no computation, while the Multiplier method incurs a marginally higher decoding time. The decoding time of both methods, however, remains competitive with that of the VTM full resolution anchor. In both cases the adapted model keeps the same architecture as the Base SR, so the additional cost is confined almost entirely to the encoder side, making the approach well suited to asymmetric encode-once, decode-many scenarios. Furthermore, at $\approx 100\,\%$ of the anchor's encoding time, both adaptive methods already yield significant luma savings, with LoRA leading, as shown in Table~\ref{tab:main_results}. LoRA also achieves clear chroma gains (U: $-14.08\%$/$-3.36\%$; V: $-22.88\%$/$-20.54\%$ for RA/AI), in contrast to RPR and Base SR, which both lose chroma. Furthermore, when measured against the non-adapted Base~SR (rather than VTM), LoRA ($r{=}4$) delivers an additional $-10.16\%$ (Y), $-39.17\%$ (U), $-39.79\%$ (V) under RA, confirming that the adaptation gain is substantial even on top of the already-beneficial SR post-filter.

\begin{figure}[t]
    \centering
    \includegraphics[width=0.85\linewidth]{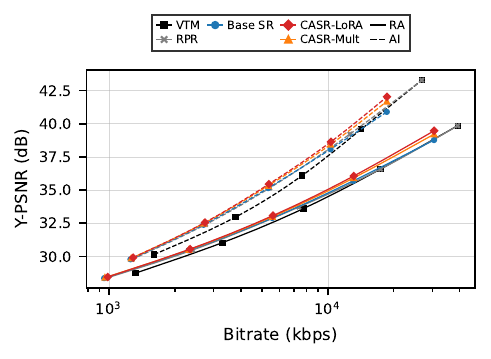}
    \caption{RD curves (Y) for ParkRunning3 (A2). Content-adaptive variants outperform SR base; LoRA yields the best trade-off.}
    \label{fig:rd_curves}
\end{figure}
\begin{figure}[t]
\centering
\setlength{\tabcolsep}{0.5pt}
\renewcommand{\arraystretch}{0.5}
\scriptsize
\begin{tabular}{cccccc}
 & Original & VTM (QP42) & SR (QP37) & +Mult (QP37) & +LoRA (QP37) \\[2pt]
\rotatebox{90}{\scriptsize\quad YUV} &
  \includegraphics[width=0.18\columnwidth]{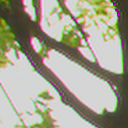} &
  \includegraphics[width=0.18\columnwidth]{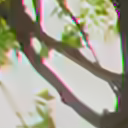} &
  \includegraphics[width=0.18\columnwidth]{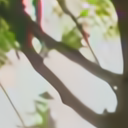} &
  \includegraphics[width=0.18\columnwidth]{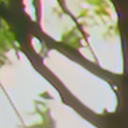} &
  \includegraphics[width=0.18\columnwidth]{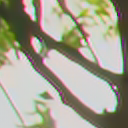} \\[1pt]
\rotatebox{90}{\scriptsize\quad Y} &
  \includegraphics[width=0.18\columnwidth]{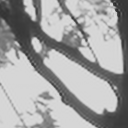} &
  \includegraphics[width=0.18\columnwidth]{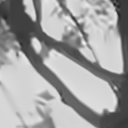} &
  \includegraphics[width=0.18\columnwidth]{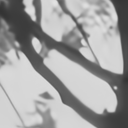} &
  \includegraphics[width=0.18\columnwidth]{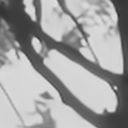} &
  \includegraphics[width=0.18\columnwidth]{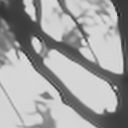} \\
 & & \tiny PSNR=29.67\,dB & \tiny PSNR=29.79\,dB & \tiny PSNR=30.33\,dB & \tiny PSNR=30.96\,dB \\[3pt]
\rotatebox{90}{\scriptsize\quad U} &
  \includegraphics[width=0.18\columnwidth]{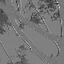} &
  \includegraphics[width=0.18\columnwidth]{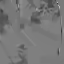} &
  \includegraphics[width=0.18\columnwidth]{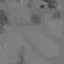} &
  \includegraphics[width=0.18\columnwidth]{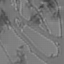} &
  \includegraphics[width=0.18\columnwidth]{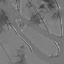} \\
 & & \tiny PSNR=27.12\,dB & \tiny PSNR=25.77\,dB & \tiny PSNR=26.35\,dB & \tiny PSNR=27.52\,dB \\[3pt]
\rotatebox{90}{\scriptsize\quad V} &
  \includegraphics[width=0.18\columnwidth]{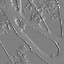} &
  \includegraphics[width=0.18\columnwidth]{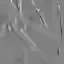} &
  \includegraphics[width=0.18\columnwidth]{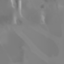} &
  \includegraphics[width=0.18\columnwidth]{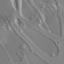} &
  \includegraphics[width=0.18\columnwidth]{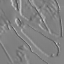} \\
 & & \tiny PSNR=26.36\,dB & \tiny PSNR=24.97\,dB & \tiny PSNR=26.14\,dB & \tiny PSNR=28.40\,dB \\
\end{tabular}
\caption{$128{\times}128$ crop from \textit{ParkRunning3} (class A2, AI). LoRA preserves fine texture and colour fidelity more faithfully than base SR or Multiplier adaptation.}
\label{fig:visual_crops}
\end{figure}

Figure~\ref{fig:rd_curves} shows rate--distortion curves for ParkRunning3. At low bitrates (high QPs), the SR post-filter consistently outperforms the native-resolution anchor because encoding at half resolution requires significantly fewer bits. At high bitrates (low QPs), the gap narrows as the native-resolution encoder preserves more detail. The LoRA-adapted post-filter shifts the RD curve upward across the entire operating range compared to the non-adapted SR baseline, demonstrating that content adaptation improves quality at every rate point.

Figure~\ref{fig:visual_crops} presents visual comparisons at a matched bitrate. The VTM anchor is encoded at full resolution with QP42, while the proposed SR methods encode at half resolution with a QP37 that produces a comparable bitrate. The luma crops show that the non-adapted SR post-filter already reconstructs the Y component well, but fails to reconstruct the chroma components faithfully. Both overfitting methods successfully restore UV components in fine detail. In particular, the LoRA-adapted model further reduces degradation in the  texturally complex region compared to the Multiplier method. The chroma crops (U, V) confirm that the improvement extends to colour fidelity, not just to luma sharpness.

\subsection{LoRA rank and signalling overhead}
\begin{table}[t]
\caption{Effect of LoRA rank on BD-rate (\%), trainable parameters, and NNR signalling overhead (OH)
compared to native-resolution VTM, under RA and AI.}
\label{tab:rank_ablation_vtm}
\centering
\footnotesize
\setlength{\tabcolsep}{3pt}
\renewcommand{\arraystretch}{0.8}
\resizebox{\columnwidth}{!}{%
\begin{tabular}{rr|rrrr|rrrr}
\toprule
            &           & \multicolumn{4}{c|}{RA}                 & \multicolumn{4}{c}{AI} \\
Rank $r$    & \% params & Y (\%)  & U (\%)  & V (\%)  & OH (\%)   & Y (\%)  & U (\%)  & V (\%)  & OH (\%) \\
\midrule
1           & $3.9\%$     & -10.13  &	-8.13   & -14.07  & 1.52      & -12.85  & 2.05    & -13.03  & 1.25 \\
2           & $7.8\%$     & -10.69  & -13.22  & -21.52 & 2.03      & -13.26     & -3.34     & -20.37     & 1.69 \\
4           & $15.6\%$    & \textbf{-10.93}  & \textbf{-15.39}  & \textbf{-24.41} & 2.96     & \textbf{-13.43}  & \textbf{-5.75}   & \textbf{-22.94}  & 2.53 \\
8           & $31.2\%$    & -10.23     & -16.75     & -26.33     & 4.77      & -12.89     & -7.00     & -24.75  & 4.16 \\
\midrule
Multiplier  & $2.1\%$     & -9.50   & -3.40   & -6.76   & 0.62      & -12.26  & 7.77    & -6.76   & 0.52 \\
\bottomrule
\end{tabular}}
\end{table}


Table~\ref{tab:rank_ablation_vtm} studies the effect of LoRA rank $r$ on coding efficiency and parameter budget. Even at $r{=}1$, LoRA already outperforms the Multiplier baseline by capturing cross-channel interactions that simple scaling cannot express. Increasing the rank to $r{=}4$ steadily improves the BD-rate, while $r{=}8$ brings diminishing returns at the cost of a larger NNR payload. The cost of these adaptations is reported in Table~\ref{tab:nnr_overhead}. The NNR-coded LoRA ($r{=}4$) payload averages approximately $41$\,kbps across all QPs, while the Multiplier payload averages approximately $9$\,kbps. Since the absolute overhead is nearly QP-independent, the relative percentage grows at higher QPs, where the base SR bitstream is smaller. At all tested QPs, this overhead remains well below the improvement in BD-rate provided by content adaptation (Table~\ref{tab:main_results}), confirming that LoRA adaptation pays for its own signalling cost. Balancing rank-driven gains against payload size, we select $r{=}4$ as the default operating point for our experiments.

\begin{table}[t]
\caption{Geometric-mean bitrate overhead (\%) from NNR-coded overfitting per QP.}
\label{tab:nnr_overhead}
\centering
\footnotesize
\setlength{\tabcolsep}{3pt}
\renewcommand{\arraystretch}{0.7}
\begin{tabular}{ll|rrrrr|r}
\toprule
Method                  & Config    & QP22       & QP27       & QP32       & QP37       & QP42       & Overall \\
\midrule
\multirow{2}{*}{LoRA($r{=}1$)}   & RA        & 0.27\%    & 0.52\%    & 1.02\%    & 1.96\%    & 3.89\%    & 1.52\% \\
                        & AI        & 0.29\%    & 0.53\%    & 0.91\%    & 1.59\%    & 2.93\%    & 1.25\% \\
\midrule
\multirow{2}{*}{LoRA($r{=}2$)}   & RA        & 0.35\%    & 0.72\%    & 1.43\%    & 2.65\%    & 5.07\%    & 2.03\% \\
                        & AI        & 0.40\%    & 0.71\%    & 1.27\%    & 2.21\%    & 3.91\%    & 1.69\% \\
\midrule
\multirow{2}{*}{LoRA($r{=}4$)}   & RA        & 0.51\%    & 1.12\%    & 2.00\%    & 3.81\%    & 7.49\%    & 2.96\% \\
                        & AI        & 0.57\%    & 1.08\%    & 1.81\%    & 3.23\%    & 6.06\%    & 2.53\% \\
\midrule
\multirow{2}{*}{LoRA($r{=}8$)}   & RA        & 0.89\%      & 1.71\%      & 3.60\%      & 6.44\%      & 11.55\%     & 4.77\% \\
                        & AI        & 1.04\%    & 1.76\%    & 3.14\%    & 5.49\%    & 9.56\%    & 4.16\% \\
\midrule
\multirow{2}{*}{Mult.}  & RA        & 0.12\%    & 0.22\%    & 0.42\%    & 0.83\%    & 1.54\%    & 0.62\% \\
                        & AI        & 0.12\%    & 0.23\%    & 0.38\%    & 0.66\%    & 1.19\%    & 0.52\% \\
\bottomrule
\end{tabular}
\end{table}

\subsection{Overfitting convergence}
\begin{figure}[t]
    \centering
    \includegraphics[width=\linewidth]{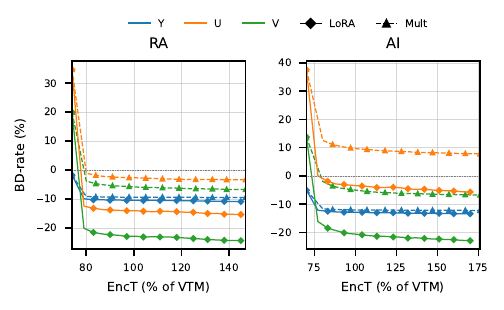}
     \caption{BD-rate (Y, U, V) vs.\ encoding time. Content adaptation progressively recovers chroma degradation from Base SR, with LoRA outperforming Multiplier across all components.}
    \label{fig:convergence}
\end{figure}

Figure~\ref{fig:convergence} traces the BD-rate evolution during overfitting. At epoch~0 (Base SR), luma already achieves moderate savings, while chroma is severely degraded. Both methods then improve all components, with most gains materialising within the first 100 epochs (at approximately 110\% encoding time for RA and 120\% for AI); ReduceOnPlateau further refines the result by lowering the learning rate upon validation-loss stagnation. LoRA consistently outperforms Multiplier throughout finetuning, most clearly on chroma.

\section{Conclusion}
\label{sec:conclusion}
We presented CASR, a content-adaptive super-resolution post-filter for VVC that is compliant with the VSEI neural network post-processing filter standard in terms of input/output tensors and in terms of including the upsampling operation within the model. The base network performs $\times 2$ upscaling internally through sub-pixel convolution. Per-sequence content adaptation is realised by overfitting Low-Rank Adaptation matrices or Multipliers attached to selected convolution layers at the encoder side; the resulting weight update is compressed with NNR and signalled as side information. Experiments on the JVET CTC class A1 and A2 sequences under both RA and AI configurations show that LoRA-based overfitting improves over both a non-adapted SR post-filter and multiplier-only overfitting, with rank $r{=}4$ providing a favourable trade-off between coding gain and signalling overhead. 

\clearpage
\balance
\bibliographystyle{IEEEbib}
\bibliography{refs}

\end{document}